\documentclass[aps,prl,twocolumn,showpacs]{revtex4-1}
\usepackage{epsfig}
\usepackage{color}
\usepackage{amsmath}
\usepackage{amssymb}
\usepackage{amsfonts}
\usepackage{bm}
\usepackage{dcolumn}
\usepackage{graphicx}
\usepackage{subfigure}
\usepackage{natbib}
\usepackage{mciteplus}
\usepackage{float}

\providecommand{\U}[1]{\protect\rule{.1in}{.1in}}

\begin{document}

\title{Radiative recombination of large polarons in halide perovskites}
\author{Mingliang Zhang$^{1,2\ast}$, Xu Zhang$^{2}$ and Hai-Qing Lin$^{1}$ }
\affiliation{$^{1}$Beijing Computational Science Research Center, Beijing 100193, China}
\affiliation{$^{2}$Department of Physics and Astronomy, California State University
Northridge, Northridge, CA 91330, USA}

\begin{abstract}
In halide perovskites, electrons (holes) exist as electronic (hole)
polarons, excitons, free and trapped electrons (holes). Six kinds of
collisions could lead to annihilation of electron and hole, three of them
involve polaron(s). In the annihilation channel of each collision process,
there is a certain probability to form a dying pair which the average
distance between electron and hole is smaller than a critical distance. The
annihilation probability per unit time of a collision process is a product
of the formation probability of the dying pair in the annihilation channel
and the annihilation probability per unit time of dying pair. To annihilate
an electronic (hole) polaron, electron (hole) must break away from the
distorted environment, which can be achieved either by tunneling or by
thermal activation. The observed temperature dependence of monomolecular and
bimolecular recombination rates, the peak frequency and line width of
photoluminescence spectrum are reproduced.
\end{abstract}

\maketitle

Organic-inorganic halide perovskites (OIHPs) have emerged as highly
promising optoelectronic materials with applications in photovoltaics \cite%
{jinsh,thom}, light-emitting diodes and low-threshold lasers \cite{sut}. All
three applications rely on the slow radiative recombination rates of
electrons and holes. In OIHPs, electrons (holes) exist as large electronic
polarons (EPs) [hole polarons (HPs)], excitons \cite{kuper}, free and
trapped electrons (holes). Simple combinations show that six kinds of binary
collisions could lead to electron and hole annihilation: (i) a free electron
and a free hole; (ii) the electron and hole in an exciton; (iii) a free
electron (hole) with a HP\ (EP); (iv) a HP and an EP; (v) a free electron
(hole) and a trapped hole (electron); and (vi) an EP (HP) and a trapped hole
(electron). To conceive new materials, one has to know the recombination
rates and statistical weights of six collisions. Measurements on mobility
\cite{mil,oga,sav,bi} indicate \cite{sm} that the majority carriers in OIHP
are EPs and HPs \cite{pro,zhu,bre,sen,ost17,mob1}. Then, the radiative
recombination involving polaron(s) [processes (iii,iv,vi)] are predominant
\cite{emi18}.

To annihilate an EP (HP), the extra electron (hole) must escape from its
distorted environment. The annihilation probability of an electron with a
hole is significant only the average distance $d_{\text{eh}}$ between
electron and hole is small (several \AA ), i.e. the electron wave function
has enough overlap with the hole wave function \cite{v4,jau}. In MAPbI$_{3}$%
, the radius $R_{\text{P}}$ of a polaron is $\thicksim $28\AA\ \cite{mob1},
which is much larger than the required `small' $d_{\text{eh}}$ for
recombination. If a free or trapped hole was at the boundary of an EP while
the extra electron was at the center of EP \cite{mitra}, the annihilation
probability would be negligible. Similarly, if an EP and a HP are in
contact, while the extra electron (hole) was at the center of EP (HP), the
annihilation probability would be negligible too. In addition, a band edge
hole (electron) cannot enter a close neighbor of the extra electron (hole)
inside an EP (HP) to annihilate. The reason is that the polarization
produced by an electron is opposite to that produced by a hole, the barrier
for an electron entering a HP\ is about two times of the polaron formation
energy ($\thicksim $140meV) \cite{mob1}. In normal operation condition, the
concentration of photo-generated electrons (holes) is less than 10$^{18}$cm$%
^{-3}$ \cite{hut}, the electron gas is non-degenerate \cite{mob1}. After
cooling, the kinetic energy of a band edge electron (hole) is $\thicksim 3k_{%
\text{B}}T/2$, which is too low to enter a HP (EP). Therefore, to annihilate
an EP (HP), the extra electron (hole) must break away from the distorted
lattice, and move to a close neighbor of the counterpart hole (electron).
The evolution of the state of the electron + lattice + hole system is driven
by the effective Coulomb attraction $V_{\text{eh}}$ between electron and
hole, the interaction $V_{\text{fm}}$ between electron (hole) and radiation
field, and the interaction $h_{\text{e-LO}}$ between electron (hole) and
longitudinal optical (LO) phonons. In OIHPs, $h_{\text{e-LO}}$ is larger
than the kinetic energy of electron (hole) and the energy of LO phonon,
conventional methods are not able to trace the radiative recombination
involving polaron(s).

In this letter, we present a tractable scheme based on three observations:
(1) If the average distance $d_{\text{eh}}$ between an electron and a hole
is smaller than a critical distance $L$, the probabilities of such an
electron-hole pair to dissolve into EP and HP, to be disassembled by thermal
excitation, to become an exciton are small, while the annihilation
probability is significant (a dying pair); (2) The extra electron (hole) in
an EP (HP) can escape the distorted lattice either through tunneling or
through thermal activation; (3) In the annihilation channel of each
collision process, there is a certain probability to form a dying pair \cite%
{sm}. The annihilation probability per unit time (APPUT) of a collision
process is a product of the formation probability of the dying pair in that
process and the APPUT of the corresponding dying pair. Had we known the
statistical weights of six kinds of collisions, the monomolecular
annihilation rate $k_{1}$ is a weighted average of the APPUT of processes
(ii,v,vi); the bimolecular annihilation rate $k_{2}$ is a weighted average
of the APPUT of processes (i,iii,iv).

We show that two types of dying pair can be formed in OIHPs: (1) both
electron and hole are movable (mobile dying pair); (2) electron (hole) is
movable and hole (electron) is trapped by a trapping center (immobile dying
pair). If $d_{\text{eh}}$ is order of or smaller than the lattice constant $%
a $, then there are a few ions between the electron and hole. The screening
caused by the displacements of ions is negligible. Then, $V_{\text{eh}}$
relates to the bare interaction by $V_{\text{eh}}=V_{\text{eh}}^{\text{bare}%
}/\varepsilon _{\infty }$, where $\varepsilon _{\infty }$ is the dielectric
constant originated from the bound electrons. For an electron-hole pair with
$d_{\text{eh}}\lesssim a$, $V_{\text{eh}}$ is the same order as the
electron-nucleus interaction, the effective masses of electron and hole are
the same as the mass $m$ of a bare electron. If both electron and hole are
mobile, the reduced mass $m_{\text{r}}^{\text{m}}$ of the pair is $m_{\text{r%
}}^{\text{m}}=m/2$; the binding energy $B_{\text{m}}$ of a mobile pair is $%
B_{\text{m}}=\varepsilon _{\infty }^{-2}m(Ke^{2})^{2}/4\hbar ^{2}$, where $%
K=(4\pi \epsilon _{0})^{-1}$; the Bohr radius of the pair is $L_{\text{m}%
}=2\varepsilon _{\infty }\hbar ^{2}/(mKe^{2})$. Similarly, for a free
electron (hole) and a trapped hole (electron), the reduced mass $m_{\text{r}%
}^{\text{i}}$ of an immobile pair is $m_{\text{r}}^{\text{i}}=m$, the
binding energy $B_{\text{i}}$ is $B_{\text{i}}=2B_{\text{m}}$, the Bohr
radius is $L_{\text{i}}=L_{\text{m}}/2$. In MAPbI$_{3}$, $\varepsilon
_{\infty }=6.5$ \cite{lin}, then $B_{\text{m}}$ $=162$meV, $L_{\text{m}}=6.8$%
\AA , $B_{\text{i}}$ $=324$meV, $L_{\text{i}}=3.4$\AA . One can see that $%
B_{m}$ and $B_{i}$ are larger than the binding energy $E_{\text{b}}$ of an
exciton (16-50meV \cite{miy}), the thermal energy (300K=26meV) and the sum
of formation free energies $F_{\text{P}}(T)$ of an EP and a HP (80-140meV)
\cite{mob1}. Therefore, an electron-hole pair with $d_{\text{eh}}<L_{m}$ ($%
L_{i}$) cannot dissolve into EP and HP, cannot be disassembled by thermal
excitation, cannot become an exciton; its fate is annihilation. We should
emphasize that a dying pair indicates all $d_{\text{eh}}<L$ configurations
not just the $d_{\text{eh}}=L$ one \cite{sm}. In the annihilation channel of
collision processes (i,ii,iii,iv), a mobile dying pair is formed; in the
annihilation channel of process (v,vi), an immobile dying pair is formed.

We first calculate the APPUT $w_{\mathrm{2f}}$ for a free electron and a
free hole. By approximating their wave-functions with plane-waves, we can
derive $w_{\text{2f}}$ based on the second-order perturbation theory with
the electron-phonon interaction and $V_{\text{fm}}$ treated as perturbations
\cite{besa,dum,schl}:
\begin{equation}
w_{\text{2f}}=\frac{e^{2}}{V4\pi \epsilon _{0}}\frac{n_{\text{cell}}\hbar
^{2}\omega _{\mathbf{k}}}{m^{2}c^{3}\omega _{\mathbf{q}}^{g}}|\frac{\epsilon
_{\mathbf{k}\sigma \beta }k_{3\beta }}{\varepsilon (\omega _{\mathbf{k}})}%
|^{2}  \label{2f}
\end{equation}%
\begin{equation*}
\frac{n_{\mathbf{q}}^{g}}{[E_{c\mathbf{k}+\mathbf{k}_{3}}-E_{c\mathbf{k}%
_{1}}-\hbar \omega _{\mathbf{q}}^{g}]^{2}}|\frac{e^{-i\mathbf{q}\cdot
\mathbf{s}_{\kappa }}}{\sqrt{M_{\kappa }}}\frac{e_{\kappa \alpha }^{g}(%
\mathbf{q})q_{\alpha }z_{\kappa }e^{2}}{\epsilon _{0}q^{2}\varepsilon
(\omega _{\mathbf{q}}^{g},T)}|^{2},
\end{equation*}%
where the repeated indices are summed over; $V$ is the volume of the sample;
$n_{\text{cell}}$ is the number of primitive cells per volume; $M_{\kappa }$
and $z_{\kappa }$ are the mass and effective nuclear charge of the $\kappa $%
th atomic core. $\mathbf{s}_{\kappa }$ is the position vector of the $\kappa
$th atom relative to the center of the primitive cell. $\mathbf{k}$ and $%
\omega _{\mathbf{k}}$ are the wave vector and frequency of emitted photon. $%
\varepsilon (\omega _{\mathbf{k}})$ is the dielectric constant at frequency $%
\omega _{\mathbf{k}}$. $\epsilon _{\mathbf{k}\sigma \beta }$ is the $\beta $%
th Cartesian component of the $\sigma $th polarization vector of photon. $%
\mathbf{k}_{1}$ is the electron wave vector in the conduction band, and $E_{c%
\mathbf{k}_{1}}$ is the energy of the electron in a state $|c\mathbf{k}%
_{1}\rangle $ of the conduction band $c$. $\mathbf{k}_{3}$ is the wave
vector of the hole in the valence band. $\mathbf{q}=\mathbf{k}+\mathbf{k}%
_{3}-\mathbf{k}_{1}$ is the phonon wave vector and $g$ is the phonon branch
index. $\omega _{\mathbf{q}}^{g}$ and $n_{\mathbf{q}}^{g}$ are the frequency
and occupation number of the phonon in mode $|g\mathbf{q}\rangle $. $w_{%
\text{2f}}$ should be understood as a sum over $\sigma $, an integration
over the direction of $\mathbf{k}$, and an average over the initial states
of electron.

The transition amplitude of a radiative recombination exponentially
decreases with the increase of $d_{\text{eh}}$ \cite{v4,jau}. Then, in the
collisions of a free electron and a free hole, the annihilation primarily
happens for those wave-packets with $d_{\text{eh}}<L_{\text{m}}$, i.e.
through a mobile dying pair. In other words, in the annihilation changel of
the free electron-free hole collision, the formation probability of a mobile
dying pair almost equals one, $w_{2f}$ approximately equals the APPUT of a
mobile dying pair. Later on, we take the APPUT of a mobile dying pair as $%
w_{2f}$.

For the electron and hole in an exciton, the formation probability of mobile
dying pair is $V|\psi (0)|^{2}$, where $\psi (0)$ represents the
wave-function of the electron at the position of the hole \cite{v4,jau}.
According to the Hydrogenic model of excitons, $|\psi (0)|^{2}=(\pi r_{\text{%
ex}}^{3})^{-1}$, where $r_{\text{ex}}=\hbar (2m_{\text{ex}}E_{\text{b}%
})^{-1/2}$ is the radius of the exciton; $m_{\text{ex}}$ is the reduced mass
of the electron and hole pair. Thus, the APPUT of an exciton is $w_{\text{ex}%
}$ $=Vw_{\text{2f}}/(\pi r_{\text{ex}}^{3})$.

To annihilate an EP (HP) with a free hole (electron), the electron (hole) of
EP (HP) must first break free from the surrounding lattice, facilitated by
thermal activation or quantum tunneling. The tunneling probability is the
greatest if the free hole is in contact with the EP. In this case, the
Coulomb attraction between them is $E_{\text{con}}^{\prime }=Ke^{2}[R_{\text{%
P}}\varepsilon (0,T)]^{-1}$, where $\varepsilon (0,T)$ is the static
dielectric function at temperature $T$. The probability that an EP and a
free hole are in contact is $p_{\text{con}}^{\prime }=(e^{E_{\text{con}%
}^{\prime }/k_{B}T}-1)/(e^{E_{\text{con}}^{\prime }/k_{B}T}+1)$. By means of
the Molecular Orbital theory, the probability $P_{\text{tun}}^{\prime }$
that the extra electron in EP tunnels to a point which its distance to the
contacted hole is $L_{\text{m}}$ is:
\begin{equation}
P_{\text{tun}}^{\prime }=\frac{R_{\text{P}}}{L_{\text{m}}}[\frac{2Ke^{2}}{R_{%
\text{P}}(B_{\text{m}}-E_{\text{P}})\varepsilon }]^{2}e^{-R_{\text{P}%
}/L_{m}},  \label{ptpe}
\end{equation}%
where $E_{\text{P}}$ is the formation energy of the polaron \cite{mob1}.
Hence the formation probability of the mobile dying pair via tunneling is $%
P_{\text{tun}}^{\prime }p_{\text{con}}^{\prime }$. The same electron can
also escape from the surrounding lattice distortion via thermal activation,
and the formation probability of the dying pair by thermal activation is $%
e^{-F_{\text{P}}/k_{B}T}$. The formation probability of the mobile dying
pair in EP-free hole collision is: $P_{\text{tun}}^{\prime }p_{\text{con}%
}^{\prime }+e^{-F_{\text{P}}/k_{B}T}$. Finally, the APPUT $w_{\mathrm{Pf}}$
for free electron (hole)-HP (EP) collision is:
\begin{equation}
w_{\text{Pf}}=[P_{\text{tun}}^{\prime }p_{\text{con}}^{\prime }+e^{-F_{\text{%
P}}/k_{B}T}]w_{\text{2f}}.  \label{kpf}
\end{equation}%
The formation probability of the mobile dying pair in the EP-HP collision
can be found similarly. The attraction energy $E_{\text{con}}$ of an EP with
a close contacted HP is $E_{\text{con}}=Ke^{2}[2R_{\text{P}}\varepsilon
(0,T)]^{-1}$. At temperature $T$, the probability that EP and HP is in
contact is $p_{\text{con}}=(e^{E_{\text{con}}/k_{B}T}-1)/(e^{E_{\text{con}%
}/k_{B}T}+1)$. Under the influence of $V_{\text{eh}}$, the electron in EP
can tunnel to a close neighbor of the hole in HP and form a mobile dying
pair. The probability $P_{\text{tun}}$ that electron tunnels into a HP and
forms a mobile dying pair is:
\begin{equation}
P_{\text{tun}}=\frac{4e^{-2R_{\text{P}}/L_{\text{m}}}}{L_{\text{m}}R_{\text{P%
}}}\frac{(Ke^{2}/\varepsilon )^{2}}{[B_{\text{m}}-E_{\text{P}}]^{2}}.
\label{pt2p}
\end{equation}%
The formation probability of mobile dying pair through thermal activation is
given by $e^{-F_{\text{P}}/k_{B}T}e^{-F_{\text{P}}/k_{B}T}$. Thus the APPUT $%
w_{\text{2P}}$ in an EP-HP collision is:
\begin{equation}
w_{\text{2P}}=[P_{\text{tun}}p_{\text{con}}+e^{-2F_{\text{P}}/k_{B}T}]w_{%
\text{2f}}.  \label{k2p}
\end{equation}

Let us consider the annihilation between a free electron (hole) and a
trapped hole (electron). We approximate the wave-function of the trapped
hole as $\chi _{h}=\pi ^{-1/2}a_{0}^{-3/2}e^{-r/a_{0}}$, where $%
a_{0}=\varepsilon _{\infty }\hbar ^{2}/(mz_{\text{t}}Ke^{2})$ is the Bohr
radius of the hole, and $z_{\text{t}}$ is the effective nuclear charge of
the trap. We can show that the APPUT $w_{\text{ft}}$ of the free
electron-trapped hole collision is \cite{v4,jau}:
\begin{equation}
w_{\text{ft}}=\frac{\hbar \omega e^{2}}{V2\pi m^{2}c^{3}\epsilon _{0}}|\int
\frac{d^{3}k_{2}}{(2\pi )^{3}}\frac{ik_{2\beta }}{\pi ^{1/2}a_{0}^{3/2}}
\label{1tr}
\end{equation}%
\begin{equation*}
\frac{2}{a_{0}[(\frac{1}{a_{0}})^{2}+k_{2}^{2}]^{2}}\frac{z_{\text{t}}e^{2}}{%
(E_{c\mathbf{k}_{2}}-E_{c\mathbf{k}_{1}})\varepsilon _{\infty }\epsilon _{0}|%
\mathbf{k}_{2}-\mathbf{k}_{1}|^{2}}|^{2},
\end{equation*}%
where $\mathbf{k}_{1}$ is the wave vector of the free electron, and $\omega $
is the photon frequency. Here $w_{\text{ft}}$ should be understood as an
average over various initial states on the right hand side of Eq.(\ref{1tr}%
). In the collision of a free electron (hole) and a trapped hole (electron),
the annihilation mainly comes from those wave-packets with $d_{\text{eh}}<L_{%
\text{i}}$, i.e. an immobile dying pair. To put it another way, in the
annihilation channel of the free electron (hole)-trapped hole (electron)
collision, the formation probability of immobile almost equals one, the
APPUT of an immobile dying pair approximately equals $w_{\text{ft}}$.

We consider the collision between an EP (HP) and a trapped hole (electron).
Because the overall charge of trapped hole (electron) and trapping center is
neutral, there is no attraction between the trapped hole (electron) and the
EP (HP). Thus, the extra electron (hole) cannot escape from EP (HP) through
tunneling, and the escape can only occur by thermal activation. Therefore,
the formation probability of the immobile dying pair is $e^{-F_{\text{P}}/k_{%
\text{B}}T}$, and the APPUT $w_{\text{Pt}}$ of the EP (HP)-trapped hole
(electron) is
\begin{equation}
w_{\text{Pt}}=e^{-F_{\text{P}}/k_{\text{B}}T}w_{\text{ft}}.  \label{kpt}
\end{equation}

Let us calculate the statistical weight of each collision. Since radiative
recombination is slower than the dissociation of excitons and polarons, we
assume that electrons (holes), excitons and EPs (HPs) are in thermal
equilibrium with each other. If exciton and polaron were not able to broken
by thermal energy, the fraction of free carriers, excitons, and polarons
would be $f_{\text{f}}^{0}=[1+e^{E_{\text{b}}/k_{\text{B}}T}+e^{F_{\text{P}%
}/k_{\text{B}}T}]^{-1}$, $f_{\text{ex}}^{0}=e^{E_{\text{b}}/k_{\text{B}}T}f_{%
\text{f}}^{0}$, and $f_{\text{P}}^{0}=e^{F_{\text{P}}/k_{\text{B}}T}f_{\text{%
f}}^{0}$, respectively. However, exciton and polaron can be broken by
thermal energy. Therefore, the percentages of electrons (or holes), excitons
and EPs (HPs) are $p_{\text{f}}=f_{\text{f}}^{0}+f_{\text{ex}}^{0}e^{-E_{%
\text{b}}/k_{B}T}+f_{\text{P}}^{0}e^{-F_{\text{P}}/k_{\text{B}}T}$, $p_{%
\text{ex}}=f_{\text{ex}}^{0}(1-e^{-E_{\text{b}}/k_{\text{B}}T})$ and $p_{%
\text{P}}=f_{\text{P}}^{0}(1-e^{-F_{\text{P}}/k_{\text{B}}T})$. Therefore,
the statistical weight of the four collisions concerning the mobile dying
pairs is: $p_{\text{2f}}=p_{\text{f}}^{2}$, $p_{\text{ex}}$, $p_{\text{Pf}%
}=p_{\text{f}}p_{\text{P}}$, and $p_{\text{2P}}=p_{\text{P}}^{2}$. Let $E_{%
\text{tra}}$ be the trap energy defined relative to the edge of the valence
(conduction) band for the hole (electron) \cite{yin15}, then the probability
that a carrier is trapped is $1-e^{-E_{\text{tra}}/k_{\text{B}}T}$. Thus the
statistical weight of free electron (hole)-trapped hole (electron) collision
is $p_{\text{ft}}=p_{\text{f}}(1-e^{-E_{\text{tra}}/k_{\text{B}}T})$, the
statistical weight of EP (HP)-trapped hole (electron) collision is $p_{\text{%
Pt}}=p_{\text{P}}(1-e^{-E_{\text{tra}}/k_{\text{B}}T})$.

The 1-body annihilation comes from processes (ii,v,vi). Hence the
monomolecular recombination rate $k_{1}$ is given by:

\begin{equation}
k_{1}(T)=p_{\text{ex}}w_{\text{ex}}+2p_{\text{ft}}(Vn_{\text{t}}w_{\text{ft}%
})+2p_{\text{Pt}}(Vn_{\text{t}}w_{\text{Pt}}),  \label{k1}
\end{equation}%
where $n_{\text{t}}$ is the density of the traps. Similarly, 2-body
annihilation comes from processes (i,iii,iv). Then the bimolecular
recombination rate $k_{2}$ is read as%
\begin{equation}
k_{2}(T)=p_{\text{2P}}(Vw_{\text{2P}})+2p_{\text{Pf}}(Vw_{\text{Pf}})+p_{%
\text{2f}}(Vw_{\text{2f}}).  \label{k2t}
\end{equation}%
The slow radiative recombination rate is caused by the small formation
probability of dying pairs in the collisions invovling polaron(s) \cite{sm}.

\begin{figure}[ht]
\centering{\includegraphics[width=0.34\textwidth]{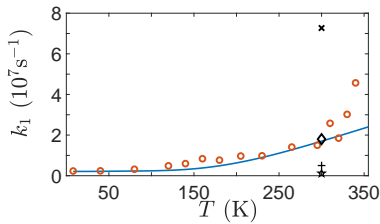}}
\caption{The monomolecular recombination rate $k_{1}$ as a function of temperature $T$ in MAPbI$_{3}$: solid curve is calculated from Eq.(\ref{k1}) and the experimental values (circles) are taken from \cite{mil},
cross from \cite{yacom}, plus symbol from \cite{rich}, diamond from \cite{yama14}, pentagon from \cite{stau}.}
\label{k1M}
\end{figure}

\begin{figure}[ht]
\centering{\includegraphics[width=0.34\textwidth]{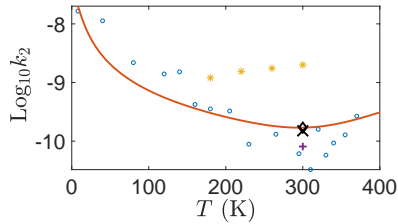}}
\caption{2-body
annihilation rate constant $k_{2}$ (cm$^{3}$s$^{-1}$) as function of
temperature T in MAPbI$_{3}$: solid line is calculated from Eq.(\ref{k2t}), experimental values circles taken from
\cite{mil}, diamond from \cite{yama14}, cross from \cite{yacom}, plus symbol from \cite{rich}, stars from \cite{hut}.}
\label{k2M}
\end{figure}

We apply Eqs.(\ref{k1},\ref{k2t}) to MAPbI$_{3}$. The materials parameters
used are: $R_{\text{P}}$ =28\AA , $E_{\text{P}}$ = 70 meV \cite{mob1}, $r_{%
\text{ex}}$ = 49 \AA , $m_{\text{ex}}=0.1m$ \cite{miy}, $z_{\text{t}}$ $=1$,
$n_{\text{t}}=3\times 10^{10}$cm$^{-3}$ \cite{shi}, $\varepsilon (0,T)$ is
taken from \cite{ono}. In Fig.\ref{k1M} and Fig.\ref{k2M} we compare the
measured $k_{1}(T)$ and $k_{2}(T)$ with Eqs.(\ref{k1},\ref{k2t}). The theory
reproduces the general experimental trends \cite{sm} that $k_{1}$ increases
monotonically while $k_{2}$ decreases first and then increases with
increasing temperature. For the three collision processes contributing to $%
k_{1}$, only the annihilation between a HP\ (EP) and a trapped electron
(hole) depends sensitively on $T$. Because the trapped hole is overall
charge neutral, there is no Coulomb attraction between the EP and the
trapped hole, thus tunneling is suppressed relative to thermal activation.
As a result, $k_{1}$ increases as temperature due to the thermal activation
of the polarons. For bimolecular recombination rate $k_{2}$, collisions
(iii,iv) depend more strongly on $T$: tunneling contribution dominates at
lower temperatures ($<310$ K) while thermal activation dominates at higher
temperatures ($>310$ K) owing to the fact that $F_{\text{P}}\thickapprox $40
- 70 meV \cite{mob1} is greater than thermal energy (300K=26meV). Below 310
K, as $T$ increases, the probability that the two polarons (or a polaron and
a free carrier) are in a close proximity necessary for tunneling is reduced,
thus $k_{2}$ decreases as $T$. Above 310 K, the thermal activation of
polarons dominates and $k_{2}$ increases as $T$.

\begin{figure}[ht]
\centering \subfigure[]{\includegraphics[width=0.23\textwidth]{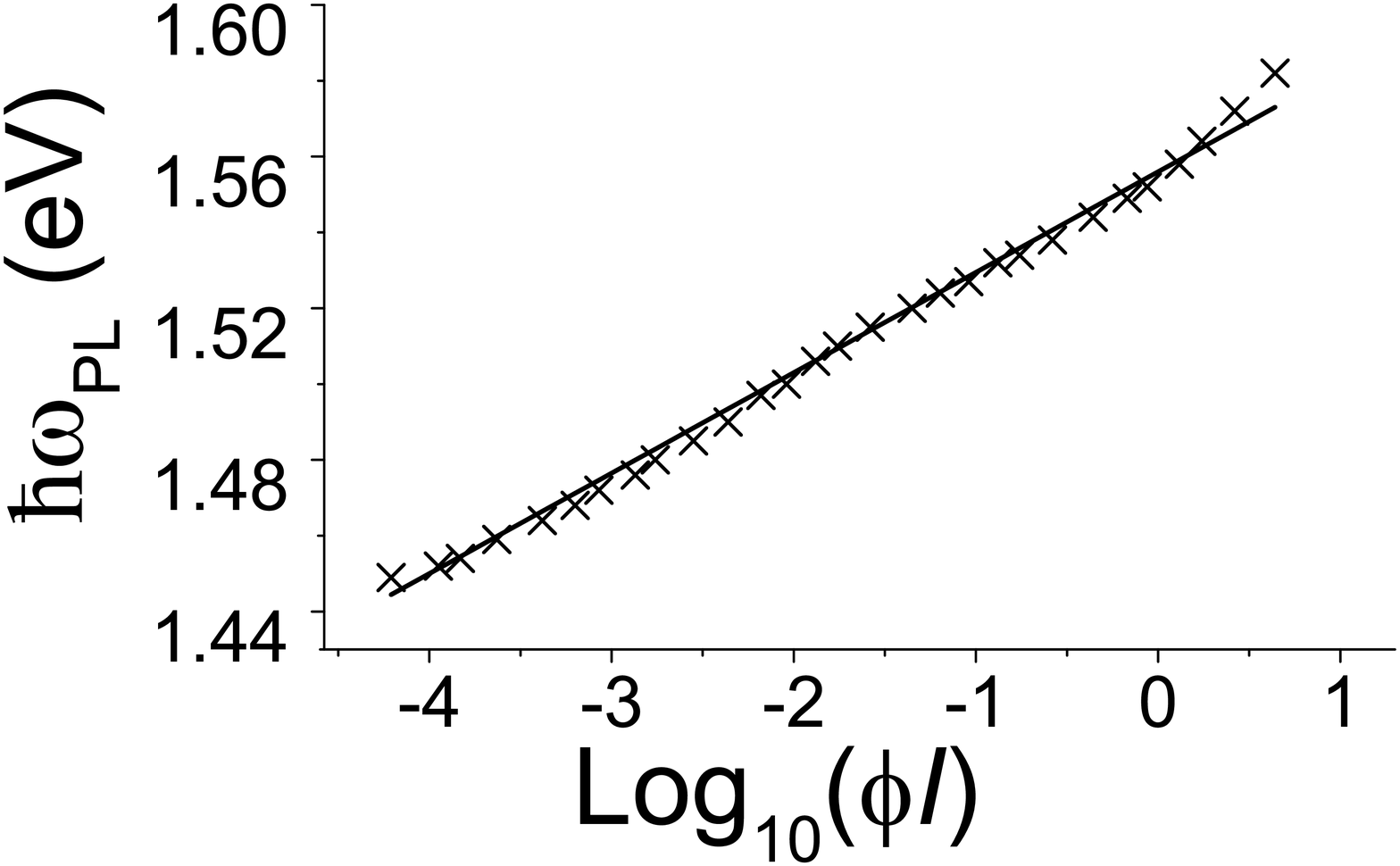}%
\label{pkf}} \subfigure[]{\includegraphics[width=0.23\textwidth]{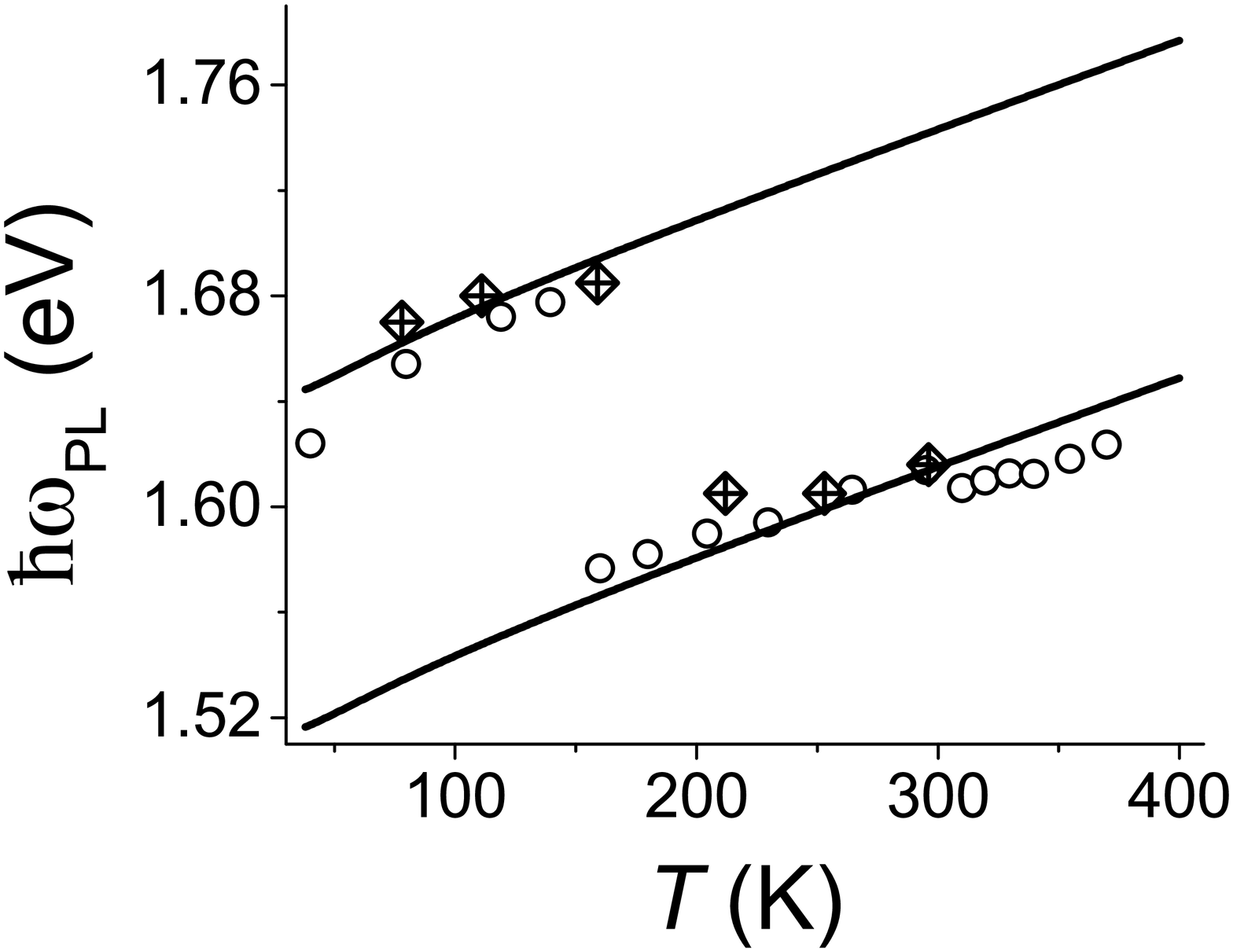}%
\label{pkM_1}}
\caption{(a) PL peak frequency $\omega_{\text{PL}}$ of MAPbI$_{3}$ as a function of the incident flux $I$: the experimental data (cross) is taken from \cite{Dare} and the solid line is
calculated from Eq.(\ref{PL}). (b) $\omega_{\text{PL}}$ as a function of temperature: the solid lines are calculated from Eq. (\ref{PL}) and the experimental values (circles) are taken from \cite{mil}, and (diamonds)
taken from \cite{eve}.} \label{pkCsM}
\end{figure}

We estimate the peak frequency $\omega _{\text{PL}}$ of PL spectrum. Since
large polarons are dominant carriers in OIHPs under normal conditions, the
PL spectrum is primarily determined by polaron recombination. $\hbar \omega
_{\text{PL}}$ approximated equals to the energy difference between the most
populated EP level and the most populated HP level. The most populated EP
level is \cite{mob1} $c_{\text{b}}+g-F_{\text{P}}(T)$, where $c_{b}$ is the
bottom of the conduction band. $g=\varepsilon _{\text{F}}[1-(\pi k_{\text{B}%
}T/2\varepsilon _{F})^{2}/3]$ is the chemical potential of polaron gas at
temperature $T$. $\varepsilon _{\text{F}}=\hbar ^{2}(3\pi
^{2}n_{e})^{2/3}/2m_{\text{P}}$ is the Fermi energy of the polaron gas. $n_{%
\text{e}}=I\phi \lbrack \varepsilon (\omega )]^{1/2}/c$ is the density of
photo-generated electrons, where $\omega $ is the excitation frequency, $%
\phi $ is the quantum yield efficiency, $c$ is the speed of light in vacuum,
and $I$ is the incident flux \cite{oga}. Similarly, the most populated HP
level is $v_{\text{t}}-g+F_{\text{P}}(T)$, with $v_{\text{t}}$ being the top
of the valence band. Therefore,
\begin{equation}
\hbar \omega _{\text{PL}}(T)=(c_{b}-v_{t})-2F_{\text{P}}(T)  \label{PL}
\end{equation}%
\begin{equation*}
+2\varepsilon _{\text{F}}[1-\frac{1}{3}(\frac{\pi k_{B}T}{2\varepsilon _{%
\text{F}}})^{2}].
\end{equation*}%
In Fig. \ref{pkf}, we plot $\omega _{\text{PL}}$ as a function of incident
light flux $I$ for MAPbI$_{3}$. The agreement with the experimental data is
very good. Furthermore, the temperature dependence of $\omega _{\text{PL}}$
\cite{sm} expected from Eq.(\ref{PL}) compare very well to the experimental
measurements \cite{mil,eve}, as shown in Fig. \ref{pkM_1}.

Although polarons are dominant carriers in OIHPs, a line width model of PL
spectrum based on free electrons (holes) \cite{leej,rud1} works well \cite%
{wri}. This contradiction can be resolved: for each collision process,
annihilation is realized via dying pair where electron and hole are no
longer confined by lattice distortion. According to Eq.(\ref{2f}), the
recombination time of the mobile dying pair is in the order of $10^{-9}$s,
which is much larger than the timescale ($\thicksim $10$^{-13}$s) of
absorbing and emitting LO phonons \cite{yey,hai}. The coupling of the
\textquotedblleft free" electron (hole) with LO phonons is the dominant
process for determining the line width \cite{sm}.

In conclusion, six kinds of binary collision could lead to radiative
recombination via dying pair. The annihilation probability per unit time of
a collision process is a product of the formation probability of the dying
pair in the annihilation channel and the annihilation probability per unit
time of the dying pair. In a recombination process involving EP, the Coulomb
attraction between the extra electron in EP and the counterpart hole helps
the extra electron in EP to escape the distorted lattice either by tunneling
or by thermal excitation. The escaped electron and counterpart hole form a
dying pair. The ansatz is applicable to all ionic and strong polar
semiconductors where large polarons are the majority of carriers.

The work at California State University Northridge was supported by the
NSF-PREM grant DMR-1205734.

$\ast $: mingliangster1@hotmail.com 
\newline
\bigskip

Supplemental Material for

\textbf{Radiative recombination of large polarons in halide perovskites}

\subsection{Polarons as the majority of carriers}

If a beam of light is shed on a halide perovskite, electrons and holes are
produced. It is well-known that in an ionic crystal, an extra electron
(hole) usually exists as large electronic (hole) polaron \cite{kuper}.
Applying theory of polaron \cite{emin} to MAPbI$_{3}$, one can estimate the
formation energy of polaron is $E_{\text{P}}$\ =70 meV, the formation free
energy $F_{\text{P}}$ is 40-70meV \cite{mob1}. The binding energy $E_{\text{b%
}}$ of an exciton is 15meV \cite{miy}. If $T<E_{\text{b}}/k_{\text{B}}$
(174K), carriers mainly exist as polarons and excitons. If $E_{\text{b}}/k_{%
\text{B}}<T<<F_{\text{P}}$, excitons eventually disappear and become
polarons. Further increasing temperature, the majority of carriers are still
polarons, the percentage of free electrons (holes) increases, $\thicksim $%
10\%.at 300K \cite{mob1}. If temperature is too high such that lattice
distortion cannot follow the motion of electron, polarons cannot exist \cite%
{dav87}.

Three behaviors of mobility $\mu $ indicate that the majority of carriers
are polarons. (1) $\mu $ depends on temperature $T$ as $\mu \propto T^{-3/2}$
\cite{oga,sav,mil}, which implies that (a) the strong 1-phonon interaction $%
h_{\text{e-LO}}$ of electron (hole) with longitudinal optical phonons does
not appear. Otherwise the temperature dependence of $\mu $ will be
different; (b) the change in distribution function is caused by the
interaction $h_{\text{e-LA}}$ of carrier with longitudinal acoustic phonons.
(2) $\mu $ is insensitive to defects \cite{zhu,bre}, which means that (c)
the carrier is massive than a bare electron, otherwise the change in wave
vector will be too large in the elastic scattering with defects; (d) the
size or de Broglie wave length of carrier is larger than a typical bond
length. (3) $\mu $ depends on the concentration $n$ of photo-generated
carriers as $\mu \propto n^{-1}$ \cite{bi,mob1}, which is possible only when
(e) the gas of carriers is non-degenerate and the number of carriers is
fixed by the incident flux. Combination of (1) and (3) requires that (f) the
effective interaction $h_{\text{e-LA}}$ is screened by a Curie-Weiss type of
dielectric function \cite{mob1}. Features (a-f) can only be explained by
assuming that the majority of carriers are large polarons \cite{mob1}.

\subsection{Dying pairs}

A dying pair means that the average distance $d_{\text{eh}}$ between
electron and hole $d_{\text{eh}}\leq L$. Denote the formation probability
for the dying pairs in the interval $(d_{\text{eh}},d_{\text{eh}}+\delta d_{%
\text{eh}})$ as $q(d_{\text{eh}})\delta d_{\text{eh}}$, one has $%
\int_{0}^{L}\delta d_{\text{eh}}\cdot q(d_{\text{eh}})=1$. Denote the
annihilation probability per unit time for pairs with $(d_{\text{eh}},d_{%
\text{eh}}+\delta d_{\text{eh}})$ as $A(d_{\text{eh}})$. Because any dying
pair with $d_{\text{eh}}\leq L$ cannot be broken by thermal activation,
become an exciton or dissolve into polaron, $A(d_{\text{eh}})$ is not very
sensitive to $d_{\text{eh}}$, $A(d_{\text{eh}})\thickapprox A(L)$. The total
annihilation probability per unit time for all pairs $0\leq $ $d_{\text{eh}%
}\leq L$ is%
\begin{equation}
\int_{0}^{L}\delta d_{\text{eh}}\cdot q(d_{\text{eh}})\cdot A(d_{\text{eh}%
})\thickapprox A(L).  \label{x3}
\end{equation}%
Eq.(\ref{x3}) means that one can use the annihilation rate for the $d_{\text{%
eh}}=L$ dying pair to approximate the total annihilation rate for all $d_{%
\text{eh}}\leq L$ dying pairs.

\subsection{Direct transition and indirect transition}

The slow charge recombination in OIHPs has been explained by the formation
of indirect band gap originated from spin-orbit coupling \cite%
{poo,etie,tiany,even,kepe} and/or lattice distortion \cite{motta,xz}. Denote
$\mathbf{q}$ as the relative shift between the bottom of the conduction band
and the top of the valence band in reciprocal space. To conserve momentum, \
one has an indirect transition. A phonon with wave vector $\mathbf{q}$ has
to be involved. For a direct band gap material, phonon assistance is not
necessary for radiative recombination. In this case, $\mathbf{q}=0$, and $w_{%
\text{2f}}$ becomes $w_{\text{2f}}^{\text{d}}$ \cite{besa} defined below:
\begin{equation}
w_{\text{2f}}^{\text{d}}=\frac{e^{2}}{V4\pi \epsilon _{0}}\frac{2\hbar
\omega _{\mathbf{k}}}{m^{2}c^{3}}|\frac{\epsilon _{\mathbf{k}\sigma \beta
}k_{3\beta }}{\varepsilon (\omega _{\mathbf{k}})}|^{2}.  \label{2f1}
\end{equation}%
Denote $R_{g\mathbf{q}}=w_{\text{2f}}/w_{\text{2f}}^{\text{d}}$. For $|%
\mathbf{q}|\ll \pi /a$, one can show that if an acoustic phonon is involved:
\begin{equation}
R\thickapprox \frac{|\mathbf{k}_{1}|^{2}n_{\text{cell}}}{|\mathbf{q}|^{5}}%
\frac{m}{M}\frac{k_{B}T}{mv_{g}^{2}}(\frac{z_{\kappa }e^{2}}{\epsilon
_{0}\hbar c_{s}})^{2},  \label{ai}
\end{equation}%
where $\mathbf{k}_{1}$, $v_{g}$, and $m$ are the wave vector, a typical
group velocity and the mass of electron, $c_{s}$ is speed of sound, $M$ is
the mass of a typical atom. If an optical phonon is involved:
\begin{equation}
R\thickapprox \frac{n_{\text{cell}}}{|\mathbf{q}|^{3}}\frac{m}{M}\frac{k_{B}T%
}{mv_{g}^{2}}(\frac{|\mathbf{k}_{1}|z_{\kappa }e^{2}}{\epsilon _{0}\hbar
\omega _{\text{LO}}})^{2}.  \label{oi}
\end{equation}

It has been shown that $|\mathbf{q}|<0.1$ \AA $\ll \pi /a$ \cite%
{poo,etie,tiany,even,kepe,motta,xz}. Using Eqs.(\ref{ai},\ref{oi}), one can
estimate that $w_{\text{2f}}/w_{\text{2f}}^{\text{d}}\thickapprox $ 0.1 -
0.3. If the charge carriers were \textquotedblleft free" electrons and holes
as opposed to large polarons, their recombination rates would be slowed down
by a factor of 3 to 10 relative to a direct band gap material, owing to the
shift of the extremes of bands in reciprocal space.

Eqs.(3,5,7) in text give the ratios of the annihilate rates involving
polaron(s) to the annihilation rates of bare electrons and holes. For MAPbI$%
_{3}$, $R_{\text{P}}$ = 28 \AA , $E_{\text{P}}$\ =70 meV, $F_{\text{P}}$
=40- 70 meV \cite{mob1}, $\varepsilon (0,T)$ is given in \cite{ono}. $L_{%
\text{m}}=6.8$\AA , $B_{\text{m}}=162$meV. Using these parameters, we find
that the formation probability of the dying pairs is in the order of 10$%
^{-3}-10^{-2}$. Therefore, the recombination rates of the polarons ($w_{%
\text{Pf}}$, $w_{\text{2P}}$, $w_{\text{Pt}}$) are 2 - 3 orders of magnitude
slower than those of bare carriers ($w_{\text{2f}}$, $w_{\text{ft}}$).
Therefore, in halide perovskites, the slow radiative recombination rate is
not caused by the small shift of band extremes, but is caused by the small
formation probability of dying pairs in the collisions involving polaron(s).

\subsection{Deviation of model from measurements}

There are apparent discrepancies between the theory and experimental values
of $k_{1}$ and $k_{2}$ around 310 K. A sharp decrease of $k_{2}$ just below
310 K may be attributed to strong ferroelectric fluctuation \cite{lius, Raki}
which could separate mobile positive and negative charges across the domains.

The 1-body annihilation is caused by three collisions: exciton, free
electron (hole)-trapped hole (electron), and EP(HP)-trapped hole (electron).
The electric field produced by ferroelectric fluctuation cannot affect the
spatial distribution of excitons and the trapped electrons (holes), because
(1) exciton is neutral; and (2) in MAPbI$_{3}$, the largest `shallow'-trap
energy $E_{\text{tra}}$ for electron is 0.192eV, the largest `shallow'-trap
energy $E_{\text{tra}}$ for hole is 0.128eV \cite{yin15}. That is why $k_{1}$
is not affected by the ferroelectric fluctuation below 310K.

The sudden rises of $k_{1}$ and $k_{2}$ at 310 K may be due to the fact that
at higher temperatures, the electrons are too fast for the lattice
deformation to follow \cite{dav87,mob1}, and thus the carriers can escape
from the surrounding lattice without resorting to tunneling or thermal
activation.

\subsection{Temperature dependence of peak frequency}

In the present work, the blue shift of $\omega _{\text{PL}}$ with increasing
temperature is attributed to the decrease of formation free energies $F_{%
\text{P}}(T)$ of EP and HP with increasing temperature \cite{mob1}. Assuming
electrons (holes) do not form EPs (HPs) but exist as free electrons (holes),
there are attempts \cite{sai16,yang17} to understand $\omega _{\text{PL}}(T)$
from the changes of CBM and VBM with $T$, a qualitative agreement with the
observed $\omega _{\text{PL}}(T)$ has been obtained. The polaron picture
does not reject a possible change in $(c_{b}-v_{t})$ with $T$ which is not
taken into account here.

\subsection{Line width of PL spectrum}

According to the general theory of line width \cite{toyo}, the line width of
PL spectrum is determined by the energy uncertainties of initial state and
final state. If the EP-HP recombination goes through activation path, the
energy uncertainty will be%
\begin{equation}
\Gamma_{1}=e^{-2F_{\text{P}}/k_{\text{B}}T}(F_{\text{P}}+k_{B}Te^{-F_{\text{P%
}}/k_{B}T}).  \label{bra}
\end{equation}
The number in bracket is less than 70meV, $e^{-2F_{\text{P}}/k_{\text{B}%
}T}<0.05$, $\Gamma_{1}<3.5$meV. If EP-HP annihilation goes through tunneling
path, the energy uncertainty is%
\begin{equation}
\Gamma_{2}=p_{\text{con}}[\frac{e^{2}}{4\pi\epsilon_{0}2R_{\text{P}%
}\varepsilon(0,T)}+\frac{\hbar^{2}}{mR_{\text{P}}^{2}}].  \label{brt}
\end{equation}
The number in bracket is less than 20meV, $p_{\text{con}}<0.4$, $%
\Gamma_{2}<8 $meV.

Since the recombination of a bare electron and a bare hole is much slower
than emitting or absorbing phonons, the energy levels of the breaking away
electron and hole are further broadened by the electron (hole)-LO phonon
interaction. In an ionic crystal, the coupling of electron (hole) with LO
phonon is strongest. The energy uncertainty is \cite{leej,rud1,wri}

\begin{equation}
\Gamma _{3}=g_{\text{LO}}n_{B}(\omega _{\text{LO}}),  \label{bf}
\end{equation}%
where
\begin{equation*}
g_{\text{LO}}\thicksim ie[\frac{\hbar \omega _{\text{LO}}}{2\epsilon
_{0}\Omega }(\frac{1}{\varepsilon _{\infty }}-\frac{1}{\varepsilon _{0}}%
)]^{1/2}\frac{\mathbf{k}\cdot \mathbf{e}_{\mathbf{k}\text{o}}}{k^{2}},
\end{equation*}%
$\Omega =a_{x}^{3}$ is volume of a primitive cell, $a_{x}$ is the length of
basis vector along x-direction, $k=|\mathbf{k}|$ is length of wave vector $%
\mathbf{k}$, $\mathbf{e}_{\mathbf{k}\text{o}}$ is polarization vector of LO
phonon \cite{cal}. Using data $\varepsilon _{0}=70$, $\varepsilon _{\infty
}=6.5$ \cite{ono}, $\hbar \omega _{\text{LO}}\thicksim 11.5$meV \cite{wri}, $%
a_{x}=6.3$\AA\ for MAPbI$_{3}$, one has $g_{\text{LO}}\thicksim 50$meV. For
most of temperature range $k_{B}T>\hbar \omega _{\text{LO}}$, then $%
n_{B}(\omega _{\text{LO}})$ is number larger than 1. Therefore, $\Gamma _{3}$
is much larger than $\Gamma _{1}$ and $\Gamma _{2}$.

\subsection{Non-radiative transition not important}

Because the deep trap centers are rare in the middle of band gap \cite{yin15}%
, the intervals between available intermediate states are much larger than $%
\hbar \omega _{\text{LO}}$, non-radiative transition by multi-phonon
emitting is improbable. By perturbation theory, one can show the probability
$w_{n}$ of a $n$-phonon emitting process per unit time is $w_{n}\thicksim
(u/a)^{n-1}w_{1}$, where $u$ is the displacement of atom, $a$ is lattice
constant, $w_{1}$ is the emitting probability per unit time for a single
phonon. For halide perovskites $w_{1}\thicksim 10^{12}$s$^{-1}$ for LO
phonon, $u/a\thicksim $ 10$^{-2}$. Then the transition probability per unit
time for a 3-phonon emitting process is $10^{6}$s$^{-1}$, which is already
slower than any radiative recombination process. The energy change in a
3-phonon process is only 50meV, while the intervals between mid-gap states
is much larger than 50meV \cite{yin15}. Therefore, non-radiative transition
is not important in halide perovskites. The nonadiabatic molecular dynamics
predicts that the life time of non-radiative transition is $\thicksim 1-$5ps
\cite{long,madj}, which is contradict to the observed long life time of
carrier (hundreds of ns) \cite{hai,mil,sav,Chen7519}.

\bibliographystyle{apsrev4-1}
\bibliography{refnew}

\end{document}